\documentclass[twocolumn,pre,showpacs,floatfix]{revtex4}
\usepackage{graphicx}
\usepackage{amsmath}
\usepackage{bm}
\usepackage{epsfig}
\usepackage{amsfonts}
\usepackage{amssymb}
\usepackage[usenames]{color}
\begin{document}

\newcommand{\brm}[1]{\bm{{\rm #1}}}
\newcommand{\Ochange}[1]{{\color{red}{#1}}}
\newcommand{\Ocomment}[1]{{\color{PineGreen}{#1}}}
\newcommand{\Hcomment}[1]{{\color{ProcessBlue}{#1}}}
\newcommand{\Hchange}[1]{{\color{BurntOrange}{#1}}}

\title{Fresh look at randomly branched polymers}
\author{Hans-Karl Janssen}
\affiliation{Institut f\"ur Theoretische Physik III, Heinrich-Heine-Universit\"at, 40225
D\"usseldorf, Germany}
\author{Olaf Stenull}
\affiliation{Department of Physics and Astronomy, University of Pennsylvania, Philadelphia
PA 19104, USA}
\date{\today}

\begin{abstract}
We develop a new, dynamical field theory of isotropic randomly branched polymers, and we use this model in conjunction with the renormalization group (RG) to study several prominent problems in the physics of these polymers. Our model provides an alternative vantage point to understand the swollen phase via dimensional reduction. We reveal a hidden Becchi-Rouet-Stora (BRS) symmetry of the model that describes the collapse ($\theta$-)transition to compact polymer-conformations, and  calculate the critical exponents to 2-loop order. It turns out that the long-standing 1-loop results for these exponents are not entirely correct. A runaway of the RG flow indicates that the so-called $\theta^{\prime}$-transition could be a fluctuation induced first order transition.
\end{abstract}
\maketitle

A single linear (non-branched) polymer in solution undergoes a second order phase transition from a swollen to a collapsed state when the solvent temperature sinks below the so-called $\theta$-point. In the swollen phase, the polymer can be thought of as a self-avoiding walk, and its radius of gyration or Flory radius scales with monomer number $N$ as $R_{N}\sim N^{\nu_{SAW}}$ ($\nu_{SAW}\geq1/2$). In the collapsed phase, the polymer assumes a compact globule-like conformation, and $R_{N}\sim N^{1/d}$ where $d$ is the dimensionality of space. The understanding of this collapse transition as a critical phenomenon has advanced considerably over the years \cite{Sch99}.

In comparison, much less is known about the collapse transition of randomly
branched polymers (RBPs). There exists a number of numerical studies
\cite{HsGr05,DeHe83,HeSe96,SeVa94,FGSW92/94,JvR97/99/00} that, taken together, indicate that the phase diagram is fairly complex including a line of collapse transitions that has qualitatively distinct parts. One part, called the $\theta$-line, corresponds to continuous
transitions with universal critical exponents of swollen RBP
configurations with mainly tree-like character to compact coil-like configurations. The other part of the transition line, called the
$\theta^{\prime}$-line, corresponds to the collapse of foam- or
sponge-like RBPs
to vesicle-like compact structures. In 2 dimensions, one finds nonuniversal exponents if this
transition is considered as continuous \cite{HsGr05}. The two different parts of
the collapse-transition line are separated by a multicritical point
which belongs to the isotropic percolation universality class. One of the open
questions is the existence of a possible further transition line between the
different configurations of collapsed RBPs. As far as theory is concerned, it is the swollen phase that is best understood mainly because the statistics of swollen RBPs can be formulated in terms of an  asymmetric Potts model~\cite{LuIs78,HaLu81,Con83} although Flory theory \cite{IsLu80} and real space renormalization \cite{FaCo80} have also been successfully applied. The former approach was used in particular to solve the field theoretic problem via a mapping of the relevant part of the asymmetric Potts model to the Yang-Lee edge problem using dimensional reduction \cite{PaSo81}. In contrast, the collapse of RBPs has been much less studied, and the current understanding mainly rests on the seminal field theoretic work of Lubensky and Isaacson (LI) \cite{LuIs78} and Harris and
Lubensky \cite{HaLu81}. However, it turns out that these papers, as far as they consider the collapse ($\theta$-)transition, contain a fundamental error in the renormalization procedure, and as a consequence the long-standing $1$-loop results for the collapse transition are strictly speaking not correct. In addition, it is not clear to date whether the $\theta^{\prime}$ transition is a second order transition or not. Therefore, we feel that the important RBP problem deserves a fresh look.

In this paper we develop a new, dynamical field theory~\cite{JaTa05} for RBPs based on a model for dynamical percolation with a tricritical instability \cite{JaMuSt04} in the non-percolating phase whose very large clusters (lattice animals), at critical values of the control parameters, have the same statistics as collapsing RBPs. We discuss the relation of our model to the asymmetric Potts model and carefully analyze its symmetries. In the swollen phase, the model has a high super-symmetry including translation and rotation invariance in super-space and leads to the well known Parisi-Sourlas dimensional reduction \cite{PaSo81}. At the collapse transition, super-rotation symmetry is broken, and we only have translation invariance in superspace, i.e., Becchi-Rouet-Stora (BRS) symmetry~\cite{BRS75}. We perform a 2-loop renormalization group (RG) calculation, that corrects and extends the long standing LI results for the collapse transition. Furthermore, we show that the $\theta^{\prime}$-transition is characterized by a runaway of the RG flow which suggests that this transition is a fluctuation induced first order transition contrary to what has been assumed in recent numerical studies~\cite{HsGr05}.

Our field theory (for background on field theory methods, we refer to~\cite{Am84,ZJ02}) is based on a generalization of the general epidemic process.  For a related approach to the somewhat simpler problem of directed randomly branched polymers, see~\cite{JWS09}. The primary fields of our theory are the field
of agents $n(\mathbf{r},t)$ and the field of the inactive debris
$m(\mathbf{r},t)=\lambda \int_{-\infty}^{t}dt^{\prime}\,n(\mathbf{r},t^{\prime})$
which ultimately forms the polymer cluster. The minimal non-Markoffian Langevin equations describing the process are given by
\begin{subequations}
\label{StochProz}%
\begin{align}
& \lambda^{-1}\partial_{t}n   =\Big[(1+cm)\nabla^{2}-r-g^{\prime}
m-\frac{f^{\prime}}{2}m^{2}\Big]n+\zeta\,,\label{Eq-Mot}\\
&\overline{\zeta(\mathbf{r},t)\zeta(\mathbf{r}^{\prime},t^{\prime})}
=\Big[\lambda^{-1}gn(\mathbf{r},t)\delta(t-t^{\prime})-fn(\mathbf{r}
,t)n(\mathbf{r^{\prime}},t^{\prime})\Big]
\nonumber \\
&\qquad\qquad\qquad\qquad\qquad\qquad\qquad\times \delta(\mathbf{r}-\mathbf{r}^{\prime
})\,. \label{abs-noise}%
\end{align}
\end{subequations}
The parameter $r$ tunes the "distance" to the percolation
threshold. Below this threshold $r$ is positive. The term proportional to $c$ describes the influence of the debris on diffusion. For the ordinary percolation problem, this term is irrelevant. As long as
$g^{\prime}>0$, the second order term $f^{\prime}m^{2}$ is
irrelevant near the transition point and the process models ordinary
percolation. We permit both signs of $g^{\prime}$ so that our model allows
for a tricritical instability. Consequently we need the second order term
$f^{\prime}>0$ for stabilization purposes, i.e., to limit the density to
finite values. The process is assumed to be locally absorbing, and thus all terms in the noise-correlation function contain at least one power of $n$. The first part of the noise correlation takes into account that the debris arises from spontaneous decay of agents, and thus $g>0$. The term proportional to $f>0$ simulates the anticorrelating behavior of the noise in regions where debris has already been produced.

Now, we refine these Langevin equations into a field theoretic model for RBPs. This procedure involves a number of nontrivial steps that we will briefly sketch in the following and that will be presented in detail elsewhere \cite{JaSt10}. As the first step, we represent the Langevin equations as a stochastic response functional
\begin{align}
\mathcal{J}&=\int d^{d}x\Big\{\lambda\int dt\tilde{n}\Big[\lambda^{-1}%
\partial_{t}-(1+cm)\nabla^{2}+r+g^{\prime}m
\nonumber \\
&+\frac{f^{\prime}}{2}m^{2}-\frac
{g}{2}\tilde{n}\Big]n+\frac{f}{2}\Big[\lambda\int dt\,\tilde{n}n\Big]^{2}%
\Big\}\,. \label{StochFu}%
\end{align}
 in the Ito-sense \cite{Ja76,DeDo76,Ja92,JaTa05}. This functional has the benefit that it allows us to systematically calculate averages $\langle \cdots \rangle$ of all sorts of observables via functional integration with weight $\exp[-\mathcal{J}]$. For studying polymers, we focus on a single cluster of a given size $N$ which we assume to emanate from a small source of strength $q$  at the origin $\mathbf{r}=0$ at time $t=0$. Then, the key quantity is the probability distribution for finding a cluster of mass $N$ given by~\cite{Ja05}
\begin{align}
\label{defP}
\mathcal{P}(N)
\sim  q \langle\delta(N-\mathcal{M})\tilde{n}(0,0)\rangle \, ,
\end{align}
where $\mathcal{M}=\int d^{d}r\,m(\mathbf{r},\infty)$. $\mathcal{P}(N)/N$ is expected to be proportional to the partition sum for interacting lattice animals \cite{HsGr05} up to an non-universal exponential factor $\sim p_0^N$ if $N$ becomes large. In actual calculations, the delta function appearing in averages like in Eq.~(\ref{defP}) is hard to handle. This problem can be simplified by averaging over Laplace-transformed observables, which are function of a variable conjugate to $N$, say $z$, and applying inverse Laplace transformation in the end. The switch to Laplace-transformed observables can be done in a pragmatic way by augmenting the original $\mathcal{J}$ with a term $z \mathcal{M}$ and then working with the new functional $\mathcal{J}_z = \mathcal{J} + z \mathcal{M}$. Because we are interested here only in the static properties of the final cluster after the epidemic has become extinct, we can greatly simplify the theory by focusing on the frequency zero part of $\mathcal{J}_{z}$, that is taking the quasistatic limit \cite{JaTa05,JaMuSt04,JaLy94} $m(\mathbf{r},t)\rightarrow
m_{\infty}(\mathbf{r})=i\tilde{\varphi}(\mathbf{r})$, $\tilde{n}%
(\mathbf{r},t)\rightarrow-i\varphi(\mathbf{r})$. Taking this limit, one has to be careful to account for the causal ordering of fields that results from the Ito calculus. In diagrammatic perturbation theory, this means that one has to rule out diagrams with closed propagator loops. An elegant way to achieve this is to use so-called ghost fields whose sole purpose is to generate additional diagrams that cancel any diagrams with non-causal loops. Such a procedure does not change the physical content of the
theory but simplifies calculations and makes it easier to find higher symmetries. The required cancellations can be achieved \cite{JaSt10} by using $D$ commuting (bosonic) fields $\chi_i$ subject to the constraint $\sum_{i=1}^D \chi_i =0$ so that they form the irreducible representation $(D,1)$ of the permutation-group $S_{D}$, and taking the limit $D \to -1$ at the end of the calculation. Furthermore, we eliminate redundant parameters by rescaling, mixing,  and shifting the fields. After all, we obtain the quasistatic
Hamiltonian
\begin{align}
\label{H_1}
\mathcal{H}  &  =\int d^{d}x\Big\{\tilde{\varphi}\bigl(\tau_{0}-\nabla
^{2}\bigr)\varphi-\frac{\tau_{1}}{2}\tilde{\varphi}^{2}+i h \tilde{\varphi
} -\frac{ig_{2}%
}{6}\tilde{\varphi}^{3}
\nonumber\\
&  +\frac{1}{2}\bigl(\tau_{0}\chi^{(2)}+(\nabla\chi)^{(2)}\bigr) +\frac{ig_{1}}{2}\tilde{\varphi}\bigl(2\tilde{\varphi}{\varphi+}\chi
^{(2)}\bigr)
\nonumber \\
&+\frac{1}{6}\bigl(3i\tilde{\varphi}(g_{0}\varphi+g_{1}^{\prime
}\tilde{\varphi})\varphi+3i(g_{0}\varphi+g_{1}^{\prime}\tilde{\varphi}%
)\chi^{(2)}
\nonumber\\
&+\sqrt{g_{0}g_{1}^{\prime}}\chi^{(3)}\bigr)\Big\}\,.%
\end{align}
where we use the shorthand notation $\chi^{(k)}=\sum_{i=1}^{D}\chi_{i}^{k}$, and where $h$ is a shifted version of the Laplace variable $z$. The $\tau$'s and the $g$'s are combinations of the original parameters, cf.\ Eqs.~(\ref{StochProz}) and (\ref{StochFu}). In particular, $\tau_0$ and $\tau_{1}$ are linearly related to $r$ and $g^{\prime}$, respectively, so that, in mean-field theory, the collapse transition corresponds to vanishing $\tau_0$, $\tau_{1}$, and $h$  and swollen RBPs correspond to vanishing $\tau_0$ and $h$, and positive and finite $\tau_1$.

What is the connection between our Hamiltonian~(\ref{H_1}) and other, established models for RBPs, Percolation and the Yang-Lee problem? To address this question, we rescale the fields so
that $g_{1}^{\prime}=g_{0}$, (which is possible, of course, only if both are non-vanishing, in particular at RG fixed points), and we define a new order parameter field with $(D+2)$ components, $s_{1}=i\tilde{\varphi}$, $s_{2}=i{\varphi}$, and for $i\geq3$: $s_{i}=\chi_{i-2}-(s_{1}+s_{2})/D$. This field satisfies the Potts constraint $\sum_{i=1}^{D+2}s_{i} =0$, and the resulting Hamiltonian with $S_{D+1}$ permutation-symmetry is that of the asymmetric $(D+2)$-state Potts model which lies at the heart of the known formulations of the RBP problem~\cite{LuIs78,HaLu81,Con83}. For $g_{1}=g_{2}=0$, the model reduces to the symmetric $(D+2)$-state Potts model with $S_{D+2}$-symmetry and thus produces the field theory of percolation in the limit $D\rightarrow-1$. For $g_{0}+2g_{1}=g_{0}+4g_{2}=0$, the Hamiltonian decomposes in a sum of $D+1$ uncoupled
Hamiltonians each describing the Yang-Lee edge problem.

To reveal the connection of our work to the results by Parisi and Sourlas~\cite{PaSo81} for swollen RBPs and to shed light on the collapse transition from a symmetry perspective, it is interesting to discuss the super-symmetries of our model. If $g_{1}^{\prime}$ is zero (or irrelevant like for finite $\tau_1>0$) non-causal loops are isolated, and can therefore be eliminated with a pair of Fermionic ghost fields $\psi$ and $\bar{\psi}$ \cite{JaLy94}. Using anticommuting super-coordinates $\theta$, $\bar{\theta
}$ with integration rules {$\int d\theta\,1={\int d\bar{\theta}\,1=}0$, $\int
d\theta\,\theta={\int d\bar{\theta}\,\bar{\theta}=}1$ and defining a
super-field $\Phi(\mathbf{r},\bar{\theta},\theta)=\varphi(\mathbf{r}%
)+\bar{\theta}\psi(\mathbf{r})+\bar{\psi}(\mathbf{r})\theta+\theta\bar{\theta
}\tilde{\varphi}(\mathbf{r})$, we can recast our model Hamiltonian as
\begin{align}
\mathcal{H}_{\mathrm{ss}}  &  =\int d^{d}xd\bar{\theta}d\theta\,\Big\{\frac
{1}{2}\Phi\bigl(\tau_{0}-\nabla^{2}-\tau_{1}\partial_{\bar{\theta}}%
\partial_{\theta}\bigr)\Phi+i\Hchange{z}\Phi\nonumber\\
& +i\Big[\frac{g_{0}}{6}\Phi^{3}+\frac{g_{1}}{2}\Phi^{2}%
(\partial_{\bar{\theta}}\partial_{\theta}\Phi)-\frac{g_{2}}{6}\Phi
(\partial_{\bar{\theta}}\partial_{\theta}\Phi)^{2}\Big]\Big\}\,.
\label{H-super1}%
\end{align}
This Hamiltonian shows BRS-symmetry
\cite{BRS75,ZJ02}, i.e., $\mathcal{H}_{\mathrm{ss}}$ is invariant under a
super-translation {$\theta\rightarrow\theta+\varepsilon,$ $\bar{\theta
}\rightarrow\bar{\theta}+\bar{\varepsilon}$. Moreover, if the control parameter
}$\tau_{1}$ is positive and finite, i.e., if we consider the problem of swollen
RBPs, $\tau_{1}$ can be reset by a scale
transformation to $2$. The super-coordinates become massive, and the
derivatives combine to a super-Laplace $\nabla^{2}+\tau_{1}\partial
_{\bar{\theta}}\partial_{\theta}\rightarrow\nabla^{2}+2\partial_{\bar{\theta}%
}\partial_{\theta}=:\square$. The coupling constants, $g_{1}$ and $g_{2}$  become irrelevant, and can be neglected. Then the Hamiltonian takes the super-Yang-Lee form and attains, besides the
super-translation invariance, super-rotation invariance. Now dimensional
reduction \cite{PaSo81} can be used to reduce the problem to the usual
Yang-Lee problem in two lesser dimensions which culminates into to well known results for swollen RBPs.

Now we come to the heart of our RG analysis, where we focus on the case that the control
parameters $\tau_{0}$ and $\tau_{1}$ take critical values (zero in mean-field
theory) where the correlation length diverges, and correlations between
different polymers vanish. The actual objects of our perturbation theory are the naively UV-divergent vertex functions $\Gamma_{\tilde{k},k}$ which consist of irreducible diagrams
with $\tilde{k}$ and $k$ amputated legs of $\tilde{\varphi}$ and $\varphi$,
respectively, as functions of the wavevector $\mathbf{q}$. We calculate these functions in dimensional regularization and $\varepsilon$-expansion about $d=6$ dimensions ($\varepsilon = 6-d$) to 2-loop order and then remove their UV divergences in minimal subtraction using the scheme
\begin{subequations}
\label{Reno}
\begin{align}
(\varphi,\tilde{\varphi},\chi)  &  \rightarrow(\mathring{\varphi}%
,\mathring{\tilde{\varphi}},\mathring{\chi})=Z^{1/2}(\varphi+K\tilde{\varphi
},\tilde{\varphi},\chi)\,,\label{Reno1}\\
\tau_i  &  \rightarrow \mathring{\tau}_i =Z^{-1}%
Z_{ij}\tau_j\,,\label{Reno2}
\\
h  &  \rightarrow \mathring{h} =Z^{-1/2}\left( h+ \textstyle{\frac{1}{2}} \mu^{-\varepsilon
/2} \tau_i A_{ij} \tau_j \right) \,,\label{Reno3}
\\
{g_{\alpha}}  &  \rightarrow\mathring{g_{\alpha}}=Z^{-3/2}(u_{\alpha
}+B_{\alpha})\mu^{\varepsilon/2}\,, \label{Reno4}%
\end{align}
\end{subequations}
where $\mu$ is an inverse length scale that is used to make the coupling constants dimensionless, where $(\tau_i )=(\tau_{0},\tau_{1})$ and $({g_{\alpha}})=(g_{0},g_{1}^{\prime},g_{1},g_{2})$. Note that the renormalization scheme introduces a counter term proportional $K$ that has no counterpart in the Hamiltonian~(\ref{H_1}). This term can be viewed as a remnant of the term proportional to $c$ in the original response functional~(\ref{StochFu}) which we removed in our journey towards $\mathcal{H}$ because $c$ is in the sense of the RG redundant. As a counter term this term is indispensable, however, because the quadratically divergent vertex function $\Gamma_{2,0}(\mathbf{q)=}\Gamma_{2,0}(\mathbf{0)+q}^{2}\Gamma_{2,0}^{\prime\prime }(\mathbf{0})+\ldots$ contains a UV-divergent $\Gamma_{2,0}^{\prime\prime}(\mathbf{0})$. This fact was overlooked by LI \cite{LuIs78} in their calculation, and their long-standing $1$-loop results are incorrect although, fortunately, the numeric deviations from the correct $1$-loop results are rather small.

As it stands, the Hamiltonian~(\ref{H_1}) has a remaining rescaling invariance that makes one of the coupling constants redundant. Before we can analyze the RG flow, we need to remove this redundancy. To this end, we switch to rescaling invariant fields $\varphi\to g_{0}^{-1}\varphi$, $\tilde{\varphi} \to g_{0}\tilde{\varphi}$, control parameters $\tau_0 \to \tau_0$, $\tau_1 \to g_{0}^{-2} \tau_1$, $h \to (2 g_0)^{-1} h$, and effective dimensionless coupling constants $u=u_{0}u_{1}^{\prime}$, $v={u_{0}u_{1}}$, $w={u_{0}^{3}}u_{2}$. The fixed points of the RG flow are determined by the zeros of the Wilson functions for the three effective couplings, $\beta_u = \mu \partial_\mu u |_0$ ($|_0$ indicates that unrenormalized quantities are kept fixed while taking derivatives) and so on. Our calculation produces
\begin{subequations}
\label{Beta-Funk}%
\begin{align}
{\beta_{u}} &  =(-\varepsilon+7u/2+10v)u\,,\label{Beta-Funk1}\\
\beta_{v} &  =(-\varepsilon+25u/6+21v/2)v-5w/6\,,\label{Beta-Funk2}\\
{\beta_{w}} &  =(-2\varepsilon+21u/2+25v)w
\nonumber\\
&-(5u^{2}+29uv/2+11v^{2})v\,,
\label{Beta-Funk3}
\end{align}
\end{subequations}
where we refrain from showing the 2-loop parts of our results due to space constraints. The picture of the RG flow that arises from these equations is the following:
The BRS-plane $u=0$ is an invariant surface of the flow equations
(\ref{Beta-Funk1}-\ref{Beta-Funk3}) to all orders and divides the
$(u,v,w)$-space in two parts: the percolation-part with $u>0$ and the
Yang-Lee-part with $u<0$ which is non-physical for the branched polymer problem. The percolation line $v=w=0$ is an invariant
line for both signs of $u$. For $u>0$ the flow goes to the percolation fixed point whereas for $u<0$ the flow tends to infinity. The Yang-Lee-line with
$a=b=0$, where $a=u+2v$ and $b=u^{2}+4w$, is also an invariant line for both
signs of $u$. For $u<0$ the flow goes to the Yang-Lee fixed point whereas
for $u>0$ the flow runs away to infinity. Altogether we have six fixed
points which are compiled in Table~\ref{tab:fixedPoint}.
Besides the trivial Gaussian fixed point we find in the BRS-plane the
stable collapse fixed point (named Collapse), and an instable fixed point (Inst2). This point lies on a separatrix in the BRS-plane and is
attracting on it. The flow of the part which contains the collapse fixed point
is of course attracting to Collapse. The other part shows runaway flow. Turning to the percolation-part of the $(u,v,w)$-space, we find one instable
point (Percolation) on the percolation line $v=w=0$. Because Percolation has two
stable directions it defines a plane that divides the space in two parts. The
flow in one of it goes to Collapse whereas the flow in the other part is again
running away. The stability plane of Percolation for $u>0$ is a continuation of
the separatrix found above on the BRS-plane for $u=0$. In the Yang-Lee-part of the $(u,v,w)$-space, we also find a plane which is the continuation of the BRS-separatrix now into the
region with $u<0$. This plane is separated in two parts by the Yang-Lee-line.
One part is attracting to an instable fixed point (Inst1), the other
part shows runaway flow. Both planes divide the $(u,v,w)$-space in a wedge-shaped part attracting to Collapse, and a part where the flow goes to infinity. The edge of the wedge is the separatrix in the BRS-plane.
\begin{table}
\begin{tabular}
[c]{|c|c|c|c|c|}\hline
& {$u_{\ast}$} & $v_{\ast}$ & $w_{\ast}$ & stability\\\hline
Gaussian & $0$ & $0$ & $0$ & $---$\\\hline
Collapse & $0$ & $\frac{\varepsilon\big(69+\sqrt{201}\big)}{760}$ & $\frac{6\varepsilon
^{2}\big(689\sqrt{201}-339\big)}{5\times760^{2}}$ & $+++$\\\hline
Percolation & $2\varepsilon/7$ & $0$ & $0$ & $++-$\\\hline
Yang-Lee & $-2\varepsilon/3$ & $\varepsilon/3$ & $-\varepsilon^{2}/9$ &
$+--$\\\hline
Inst1 & $-\varepsilon/2$ & $11\varepsilon/40$ & $-517\varepsilon^{2}/8000$ &
$++-$\\\hline
Inst2 & $0$ & $\frac{\varepsilon\big(69-\sqrt{201}\big)}{760}$ & $\frac{-6\varepsilon
^{2}\big(689\sqrt{201}+339\big)}{5\times760^{2}}$ & $+--$\\\hline
\end{tabular}
\caption{RG fixed points to leading order.}
\label{tab:fixedPoint}
\end{table}
Note that the flow diagram has the following perhaps unexpected implication for the $\theta^\prime$-transition. The region behind the percolation plane where $u$ runs away to ever more positive values indicate that this transition might be discontinuous and not, as previously assumed, a second order transition.

Finally, we compile our main results for the collapse transition. Our RG analysis leads to three independent critical exponents. For the probability distribution $\mathcal{P}(N)$, we find the asymptotic  form
\begin{equation}
\mathcal{P}(N,y)/N \sim N^{-\theta}\mu_{0}^{N}f_{\mathcal{P}} \big(yN^{\phi} \big)\,, \label{Skal-Pop}%
\end{equation}
where $\mu_{0}$ is a non-universal constant, $f_{\mathcal{P}}$ is a scaling function, and the effective control parameter for the "distance" from the transition is given by the scaling variable $y$ which is a linear combination of $\tau_0$ and $\tau_1$. For the radius of gyration, we obtain
\begin{equation}
R (N,y)\sim  N^{\nu} f_{R} \big(yN^{\phi} \big)\,. \label{Skal-Gyr}%
\end{equation}
To second order in $\varepsilon$-expansion, the critical exponents of the collapse transition read
\begin{subequations}
\begin{align}
{\theta=}  & {\frac{5}{2}-0.4925\,({\varepsilon/6})-0.5778\,({\varepsilon
/6})^{2}\,,}\\
\phi=  & \frac{1}{2}+0.0225\,({\varepsilon/6})-0.3580\,({\varepsilon/6}%
)^{2}\,,\\
{\nu=}  & {\frac{1}{4}+0.1915\,({\varepsilon/6})+0.0841\,({\varepsilon
/6})^{2}\,.}%
\end{align}
\end{subequations}
Note that these results compare well within the expectations for large $\varepsilon$ with recent simulations in $d=2$~\cite{HsGr05}.

In summary, we have presented a new renormalized field theory for RBPs. Though almost a classic physics problem, RBPs are still a lively subject of current research with important open questions some of which our work can help to settle.


\begin{thebibliography}{99}


\bibitem {Sch99}For a review, see e.g., L.\ Sch\"afer, \emph{Excluded Volume Effects
in Polymer Solutions} (Springer-Verlag, Berlin, 1999).

\bibitem {HsGr05}H.-P.\ Hsu and P.\ Grassberger, J.\ Stat.\ Mech., P06003 (2005).

\bibitem {DeHe83}B.\ Derrida and H.J.\ Herrmann, J.\ Physique \textbf{44},
1365 (1983).

\bibitem {HeSe96}M.\ Henkel and F.\ Seno, Phys.~Rev.~E \textbf{53}, 3662 (1996).

\bibitem {SeVa94}F.\ Seno and C.\ Vanderzande, J.\ Phys.\ A:
Math.\ Gen.\ \textbf{27}, 5813, 7937 (1994).

\bibitem {FGSW92/94}S.\ Flesia, D.S.\ Gaunt, C.E.\ Soteros and
S.G.\ Whittington, J.\ Phys.\ A: Math.\ Gen.\ \textbf{25}, L1169 (1992);
\textbf{27}, 5831 (1994) .

\bibitem {JvR97/99/00}E.J.\ Janse van Rensburg et al., J.\ Phys.\ A:
Math.\ Gen.\ \textbf{30}, 8035 (1997); \textbf{32}, 1567 (1999); \textbf{33},
3653 (2000).

\bibitem {LuIs78}T.C.~Lubensky and J.~Isaacson, Phys.~Rev.~Lett.~\textbf{41},
829 (1978); Phys.~Rev.~Lett.~\textbf{42}, 410(E) (1978); Phys.~Rev.~A
\textbf{20}, 2130 (1979).

\bibitem {HaLu81}A.B.~Harris and T.C.~Lubensky, Phys.~Rev.~B \textbf{23}, 3591
(1981); Phys.~Rev.~B \textbf{224}, 2656 (1981).

\bibitem {Con83}A.\ Coniglio, J.\ Phys.\ A: Math.\ Gen.\ \textbf{16}, L187 (1983).

\bibitem {IsLu80}J.~Isaacson and T.C.~Lubensky, J.\ Physique
Lett.\ \textbf{41}, L469 (1980).

\bibitem {FaCo80}F.~Family and A.~Coniglio, J.~Phys.~A: Math.~Gen.~\textbf{13}, L403 (1980).

\bibitem {PaSo81}G.\ Parisi and N.\ Sourlas, Phys.~Rev.~Lett.\ \textbf{46},
871 (1981).

\bibitem {JaSt10}H.K.~Janssen and O.~Stenull, forthcoming paper.

\bibitem {JaTa05}For a review on dynamical field theory  in the context of
percolation see H.K.~Janssen and U.C.~T\"{a}uber,
Ann.\ Phys.\ (N.Y.) \textbf{315}, 147 (2005).

\bibitem {JaMuSt04}H.K.~Janssen, M.~M\H{u}ller, and O.~Stenull, Phys.~Rev.~E
\textbf{70}, 026114 (2004).

\bibitem {Ja05}H.K.~Janssen, J.~Phys.~C: Cond.~Mat.~\textbf{17}, S1973 (2005).

\bibitem {BRS75}C.~Becchi, A.~Rouet, and R.~Stora, Comm.~Math.~Phys.~
\textbf{52}, 55 (1975).

\bibitem {Am84}D.J.\ Amit, \emph{Field Theory, the Renormalization Group, and
Critical Phenomena} (World Scientific, Singapore, 1984).

\bibitem {ZJ02}J.\ Zinn-Justin, \emph{Quantum Field Theory and Critical
Phenomena} (Clarendon, Oxford, fourth edition 2002).

\bibitem{JWS09}H.K.~Janssen, F.~Wevelsiep, and O.~Stenull, Phys.~Rev.~E~\textbf{80}, 041809 (2009).

\bibitem {Ja76}H.K.\ Janssen, Z.\ Phys.\ B~\textbf{23}, 377 (1976);
R.\ Bausch, H.K.\ Janssen, and H.\ Wagner, Z.\ Phys.\ B~\textbf{24}, 113 (1976).

\bibitem {DeDo76}C.\ DeDominicis, J.\ Physique C~\textbf{37}, 247 (1976);
C.\ DeDominicis and L.\ Peliti, Phys.\ Rev.\ B~\textbf{18}, 353 (1978).

\bibitem {Ja92}H.K.\ Janssen, in: \emph{Dynamical Critical Phenomena and
Related Topics}, Lecture Notes in Physics, Vol.\ 104, ed.\ C.P.\ Enz
(Springer, Heidelberg, 1979); H.K.\ Janssen, in: \emph{From Phase Transition
to Chaos}, ed.\ G.\ Gy\"{o}rgyi, I.\ Kondor, L.\ Sasv\'{a}ri, T.\ T\'{e}l
(World Scientific, Singapore, 1992).

\bibitem {JaLy94}H.K.~Janssen and A.~Lyssy, Phys.~Rev.~E \textbf{50}, 3784 (1994).



\end{thebibliography}
\end{document}